\documentclass[12pt,a4paper]{article}
\usepackage[utf8]{inputenc}
\usepackage[english]{babel}
\usepackage{amsmath, amssymb, amsthm}

\usepackage{graphicx}
\usepackage{braket}
\usepackage{geometry}
\usepackage{tikz}
\usetikzlibrary{quantikz2}

\newcommand{\bplus}{\pmb{+}}
\newcommand{\bminus}{\pmb{-}}
\newcommand{\bplusi}{\pmb{+i}}
\newcommand{\bminusi}{\pmb{-i}}
\DeclareMathOperator{\Tr}{Tr}
\title{Correlation-Based Measurement of Entanglement in Rank-2 Mixed States}
\author{
 M. I. Samar and V. M. Tkachuk\\
Department for Theoretical Physics,\\
Ivan Franko National University of Lviv, Ukraine\\
12, Drahomanov St., Lviv, 79005, Ukraine}
\date{\today}
\usepackage{subcaption}

\begin{document}
\maketitle

\begin{abstract}
We study correlation-based entanglement evaluation in rank-2 mixed states of two- and multi-qubit systems. The relevant two-dimensional subspace is treated as a logical qubit embedded in the physical Hilbert space, and factorizable realizations of its logical Pauli operators are classified. For two qubits, this yields nontrivial canonical encodings related by local Clifford transformations. The relation between logical and physical Bloch vectors admits a geometric interpretation as a transversely deformed Bloch sphere, leading to closed-form expressions for two-qubit concurrence and generalized concurrence across $N$-qubit bipartitions. The entanglement scaling factor is obtained directly from complementary-subsystem correlations, without prior calibration of state-preparation parameters. Noisy simulations using the IBM Quantum \texttt{FakeKingston} backend model, with readout-error mitigation and zero-noise extrapolation, confirm the analytical results up to $N=10$ qubits. The protocol therefore enables entanglement estimation from only a small number of measurement settings for this class of rank-2 logical states.
\end{abstract}
\section{Introduction}

Quantum entanglement is one of the fundamental features of quantum
mechanics \cite{Einstein1935,Schrodinger1935} and a key resource for
quantum information processing and quantum communication
\cite{Horodecki2009}. For mixed states, the distinction between
classical statistical correlations and genuinely quantum correlations
is particularly important. A bipartite state $\rho$ defined on
$\mathcal H_{AB}=\mathcal H_A\otimes\mathcal H_B$ is separable if it
can be represented as a convex combination of product states
\cite{Werner1989},
\begin{equation}
    \rho
    =
    \sum_i p_i
    \rho_A^{(i)}
    \otimes
    \rho_B^{(i)},
    \qquad
    p_i\geq0,
    \qquad
    \sum_i p_i=1.
\end{equation}
States that do not admit such a decomposition are entangled.

Identifying and quantifying entanglement without reconstructing the
complete density matrix remains an important problem
\cite{Lu2016}. Full quantum state tomography (QST) provides complete
information about $\rho$, but its measurement cost grows rapidly with
the system size. For two-qubit systems, concurrence is a notable
exception in that an exact closed-form expression is known for an
arbitrary mixed state \cite{Wootters1998}. Its evaluation, however,
generally requires sufficiently complete information about the density
matrix, which motivates approaches aimed at extracting entanglement
more directly from experimentally accessible observables.
Entanglement witnesses provide a less demanding approach to detecting
particular classes of entangled states
\cite{Terhal2000,Lewenstein2000,Guhne2009}, and measured witness
expectation values can also be used to obtain quantitative bounds on
entanglement measures \cite{Guhne2007}.

Several approaches have been proposed to access entanglement more
directly without full tomography. Multipartite concurrence of pure
states can be determined from measurements on two copies of the state
\cite{Aolita2006}, while mixed-state entanglement can be estimated
through local parity measurements on two copies
\cite{Mintert2007}. Such direct two-copy measurements have also been
demonstrated experimentally for mixed photonic states
\cite{Schmid2008}. For rank-2 two-qubit mixed states, the concurrence
can be determined exactly through local projective measurements on
multiple copies of the state \cite{Yu2008}. Collective-observable
approaches have also been developed for universal two-qubit
entanglement detection \cite{Augusiak2008}, while related bipartite
quantum correlations can be accessed experimentally from a small set
of observables measured on several copies \cite{Girolami2012}.
Fundamental limitations on deciding entanglement of completely unknown
states without acquiring tomographically complete information were
discussed in Ref.~\cite{Carmeli2016}. Other proposals include
concurrence quantification through weak measurements
\cite{Tukiainen2017} and two-copy weak-value protocols for universal
two-qubit entanglement detection \cite{Goswami2019}.

More recent approaches include randomized measurements for the
quantification of multipartite concurrence \cite{Ohnemus2023},
incomplete local measurements combined with machine-learning methods
\cite{Koutny2023}, and symmetric-measurement-based lower bounds of
concurrence that can be accessed without full state tomography
\cite{Wang2025}. Very recently, exact concurrence estimation from a
single expectation value has also been proposed for known two-qubit
target states by combining a tailored unitary transformation with
optimal control \cite{Rodriguez2026}. These developments illustrate
the continuing effort to reduce the measurement resources required
for entanglement quantification.

Restricted low-rank state families provide an important setting in
which exact and experimentally accessible relations can be obtained.
A geometric description of polynomial entanglement measures for
rank-2 states in terms of an effective Bloch sphere was developed in
Ref.~\cite{Regula2016}, providing exact results for broad classes of
low-rank states. In particular, geometric entanglement measures for
rank-2 mixed states have been related to spin expectation values and
correlations \cite{Frydryszak2017}, and correlation-based protocols
for measuring the entanglement of rank-2 mixed states on quantum
computers were studied in Ref.~\cite{Kuzmak2021}.
These results show that the two-dimensional structure of rank-2 states
can substantially simplify entanglement quantification.
A complementary question is which embeddings of such an effective
two-level system into the physical Hilbert space admit simple
factorized physical observables representing the logical Pauli
operators. This question is particularly relevant for constructing
measurement protocols based on a small number of directly accessible
correlations.

In this work, we focus on rank-2 states described within a
two-dimensional logical subspace and ask under which conditions their
entanglement can be obtained directly from simple physical
correlations. We classify factorizable realizations of the logical
Pauli operators and use this structure to construct correlation-based
expressions for concurrence. The same approach is then extended to
bipartite $N$-qubit systems. This provides a direct connection between
the logical-qubit description of rank-2 states and experimentally
accessible physical observables.

This paper is organized as follows. Section~2 introduces the
Bloch-vector description of rank-2 logical states. Section~3
classifies factorizable logical Pauli operators and constructs the
corresponding two-qubit encodings. Section~4 derives the closed-form
concurrence for two-qubit rank-2 mixed states and relates it to
directly measurable Pauli correlations. Section~5 generalizes the
construction to $N$-qubit bipartite systems and establishes its
relation to global and subsystem purities. Section~6 presents noisy
numerical simulations and error-mitigation results, while Sec.~7
summarizes the main conclusions.

\section{Rank-2 Mixed Logical States and Bloch-Vector Formalism}

We consider a mixed quantum state represented by a statistical ensemble of pure states,
\begin{equation}
    \rho=\sum_i p_i |\psi_i\rangle\langle\psi_i|,
    \qquad
    p_i\geq 0,
    \qquad
    \sum_i p_i=1,
\end{equation}
where all constituent states are restricted to the same two-dimensional logical subspace
\begin{equation}
    \mathcal{H}_{\mathcal L}
    =
    \operatorname{span}
    \left\{
    |\Uparrow\rangle,
    |\Downarrow\rangle
    \right\}.
\end{equation}
Thus,
\begin{equation}
    |\psi_i\rangle
    =
    a_i|\Uparrow\rangle+b_i|\Downarrow\rangle,
    \qquad
    |a_i|^2+|b_i|^2=1.
\end{equation}

{Since $\rho$ acts nontrivially only within the two-dimensional
logical subspace $\mathcal H_{\mathcal L}$, its rank is at most two.}
Consequently, $\rho$ can be {represented as an effective
single-qubit density matrix} by expanding it over the identity operator
$I$ and the set of logical Pauli operators
$\vec{\Sigma}=(\Sigma^x,\Sigma^y,\Sigma^z)$:
\begin{equation}
\rho = \frac{1}{2} \left( I + \vec{R} \cdot \vec{\Sigma} \right)
= \frac{1}{2}
\begin{pmatrix}
1 + R^z & R^x - i R^y \\
R^x + i R^y & 1 - R^z
\end{pmatrix}.
\end{equation}
Here, $I$ and $\Sigma^\alpha$ are understood as operators acting within the two-dimensional logical subspace $\mathcal{H}_{\mathcal L}$, and the matrix representation is written in the logical basis
{$\{|\Uparrow\rangle,|\Downarrow\rangle\}$}.

The components of the logical Bloch vector
$\vec R=(R^x,R^y,R^z)$ are therefore
\begin{equation}
    R^\alpha
    =
    \Tr(\rho\Sigma^\alpha),
    \qquad
    \alpha\in\{x,y,z\}.
\end{equation}
Each pure state $|\psi_i\rangle$ corresponds to a unit logical
Bloch vector
\begin{equation}
    \vec R_i
    =
    (R_i^x,R_i^y,R_i^z),
    \qquad
    |\vec R_i|=1,
\end{equation}
while for the mixed state
\begin{equation}
    \vec R
    =
    \sum_i p_i \vec R_i,
    \qquad
    |\vec R|\leq 1.
\end{equation}
Consequently, statistical mixing within the logical subspace has the
usual geometrical interpretation as {moving from the surface
of the logical Bloch sphere toward its interior}.

In what follows, we focus on logical Pauli operators that
{admit factorizable representations as single tensor products
of local physical observables}. This restriction is particularly useful
experimentally because the corresponding logical expectation values can be
{obtained directly from physical multi-qubit correlations}.
\section{Two-Qubit Factorizable Logical Operators and Subspace Construction}

We first consider a two-qubit system with Hilbert space
$\mathcal H=\mathcal H_1\otimes\mathcal H_2$.
A general factorizable Hermitian logical operator can be written as
\begin{equation}
    \Sigma^k=A_k\otimes B_k,
    \qquad k\in\{x,y,z\},
\end{equation}
where
\begin{equation}
    A_k=\alpha_k I_1+\vec u_k\cdot\vec\sigma_1,
    \qquad
    B_k=\beta_k I_2+\vec v_k\cdot\vec\sigma_2.
\end{equation}
Here,
\begin{equation}
    \vec\sigma_j=(\sigma_j^x,\sigma_j^y,\sigma_j^z),
    \qquad j\in\{1,2\},
\end{equation}
is the vector of Pauli operators acting on the $j$th physical qubit,
and $I_j$ is the corresponding $2\times2$ identity operator.
We choose the local factors $A_k$ and $B_k$ to be Hermitian; therefore,
$\alpha_k,\beta_k$ and the components of $\vec u_k,\vec v_k$ are real.

{Within the class considered here, the physical
representatives of the logical Pauli operators are required to satisfy
the Pauli algebra in the full two-qubit Hilbert space,}
\begin{equation}
    (\Sigma^k)^2=I,
    \qquad
    \{\Sigma^i,\Sigma^j\}=0
    \quad (i\neq j),
    \qquad
    \Sigma^x\Sigma^y=i\Sigma^z,
\end{equation}
together with the corresponding cyclic relations, where
$I=I_1\otimes I_2$ is the identity operator in the two-qubit Hilbert space.

Using
\begin{equation}
    (\vec u\cdot\vec\sigma)^2=|\vec u|^2I,
\end{equation}
{and the condition
$(A_k\otimes B_k)^2=I_1\otimes I_2$, one finds that
$A_k^2$ and $B_k^2$ must be proportional to the corresponding
identity operators. After rescaling the two tensor factors by
reciprocal constants, each factor may therefore be chosen as a
Hermitian unitary. For a single qubit, such an operator is either
$\pm I$ or $\pm\vec u\cdot\vec\sigma$, with $|\vec u|=1$.
Overall signs can be absorbed into the definition of the logical
Pauli axes.}
Thus, a factorizable logical Pauli operator is either local,
\begin{equation}
    \Sigma^k=(\vec u_k\cdot\vec\sigma_1)\otimes I_2,
\end{equation}
or genuinely two-qubit,
\begin{equation}
    \Sigma^k
    =
    (\vec u_k\cdot\vec\sigma_1)
    \otimes
    (\vec v_k\cdot\vec\sigma_2),
\end{equation}
with analogous expressions obtained by interchanging the two qubits.

{We classify factorizable Pauli triples by the number of local operators. If two logical Pauli operators are local, their anticommutation forces them to act on the same physical qubit; consequently, $\Sigma^z=-i\Sigma^x\Sigma^y$ is local on that qubit as well. The triple is therefore fully local and cannot generate bipartite entanglement, so we exclude this case.}

{A triple of three genuinely two-qubit Pauli products is likewise impossible. Two such operators can anticommute only on one physical qubit; on the other, their nontrivial Pauli factors must commute and are therefore identical up to sign. Their product then becomes local on the first qubit.}

{Hence, the only nontrivial factorizable realization consists of one local logical Pauli operator and two genuinely two-qubit operators sharing the same Pauli factor, up to sign, on the other physical qubit.}

Choosing $\Sigma^z$ as the local operator gives
\begin{equation}
\begin{aligned}
    \Sigma^x
    &=
    (\vec u_x\cdot\vec\sigma_1)
    \otimes
    (\vec v\cdot\vec\sigma_2),\\
    \Sigma^y
    &=
    (\vec u_y\cdot\vec\sigma_1)
    \otimes
    (\vec v\cdot\vec\sigma_2),\\
    \Sigma^z
    &=
    (\vec u_z\cdot\vec\sigma_1)\otimes I_2,
\end{aligned}
\end{equation}
where $\vec u_x,\vec u_y,\vec u_z$ form an orthonormal right-handed
triad and $\vec v$ is a unit vector. By suitable local unitary
transformations, these vectors can be
aligned with the standard Cartesian axes,
\begin{equation}
\vec u_x=\hat x,\qquad
\vec u_y=\hat y,\qquad
\vec u_z=\hat z,\qquad
\vec v=\hat z.
\end{equation}
This gives a convenient canonical form up to local unitary
transformations,
\begin{equation}
\Sigma^x=\sigma_1^x\otimes\sigma_2^z,\qquad
\Sigma^y=\sigma_1^y\otimes\sigma_2^z,\qquad
\Sigma^z=\sigma_1^z\otimes I_2.
\label{eq:canonical_algebra}
\end{equation}
This choice corresponds to Case~III in
Table~\ref{tab:logical_operators}, where $\Sigma^z$ is the logical
Pauli operator with a local physical realization.

Since $\Sigma^z=\sigma_1^z\otimes I_2$, the logical basis states can be
chosen as
\begin{equation}
\begin{aligned}
|\Uparrow\rangle
&=
\cos\frac{\theta}{2}|00\rangle
+
e^{i\phi}\sin\frac{\theta}{2}|01\rangle,\\
|\Downarrow\rangle
&=
\Sigma^x|\Uparrow\rangle
=
\cos\frac{\theta}{2}|10\rangle
-
e^{i\phi}\sin\frac{\theta}{2}|11\rangle.
\end{aligned}
\end{equation}
By construction,
\begin{equation}
\langle\Uparrow|\Downarrow\rangle=0,\qquad
\Sigma^z|\Uparrow\rangle=|\Uparrow\rangle,\qquad
\Sigma^z|\Downarrow\rangle=-|\Downarrow\rangle.
\end{equation}

The choice of $\Sigma^z$ as the local logical operator is not unique;
the analogous canonical representations with local $\Sigma^x$ or
$\Sigma^y$ correspond to Cases~I and II, respectively, and are
summarized in Table~\ref{tab:logical_operators}.

\begin{table}[htbp]
\centering

\renewcommand{\arraystretch}{2.1}

\begin{tabular}{|c|l|c|c|c|}
\hline

\textbf{Case}
&
\textbf{Logical Basis States}
&
$\Sigma^x$
&
$\Sigma^y$
&
$\Sigma^z$
\\[0.4ex] \hline

I
&
$\!\begin{aligned}
\ket{\Uparrow}
&=
\cos\tfrac{\theta}{2}\ket{00}
+
e^{i\phi}\sin\tfrac{\theta}{2}\ket{11}
\\
\ket{\Downarrow}
&=
\cos\tfrac{\theta}{2}\ket{10}
+
e^{i\phi}\sin\tfrac{\theta}{2}\ket{01}
\end{aligned}\!$
&
$\sigma_1^x \otimes I_2$
&
$\sigma_1^y \otimes \sigma_2^z$
&
$\sigma_1^z \otimes \sigma_2^z$
\\[0.9ex] \hline

II
&
$\!\begin{aligned}
\ket{\Uparrow}
&=
\cos\tfrac{\theta}{2}\ket{00}
+
e^{i\phi}\sin\tfrac{\theta}{2}\ket{11}
\\
\ket{\Downarrow}
&=
\cos\tfrac{\theta}{2}\ket{10}
-
e^{i\phi}\sin\tfrac{\theta}{2}\ket{01}
\end{aligned}\!$
&
$\sigma_1^x \otimes \sigma_2^z$
&
$\sigma_1^y \otimes I_2$
&
$\sigma_1^z \otimes \sigma_2^z$
\\[0.9ex] \hline

III
&
$\!\begin{aligned}
\ket{\Uparrow}
&=
\cos\tfrac{\theta}{2}\ket{00}
+
e^{i\phi}\sin\tfrac{\theta}{2}\ket{01}
\\
\ket{\Downarrow}
&=
\cos\tfrac{\theta}{2}\ket{10}
-
e^{i\phi}\sin\tfrac{\theta}{2}\ket{11}
\end{aligned}\!$
&
$\sigma_1^x \otimes \sigma_2^z$
&
$\sigma_1^y \otimes \sigma_2^z$
&
$\sigma_1^z \otimes I_2$
\\[0.9ex] \hline

\end{tabular}

\caption{Logical basis states and corresponding factorizable logical
Pauli operators for the three canonical choices.}
\label{tab:logical_operators}
\end{table}

Although the three cases correspond to different choices of the
logical Pauli operator with a local physical realization, they
represent equivalent canonical realizations of the same logical Pauli
algebra. Once the local physical axes have been brought to the
canonical Pauli directions, the corresponding operator triples can
be transformed into one another by local Clifford transformations,
combined with permutations and possible sign changes of the logical
Pauli axes.

For example, applying a Hadamard gate to the first physical qubit
transforms the Case~III operators as
\begin{equation}
    (X_1Z_2,Y_1Z_2,Z_1I_2)
    \xrightarrow{H_1}
    (Z_1Z_2,-Y_1Z_2,X_1I_2),
\end{equation}
which coincides with the Case~I representation after exchanging the
logical $x$ and $z$ axes and reversing the $y$ axis.

The basis states transform consistently according to
\begin{equation}
    H_1|\Uparrow\rangle_{\rm III}
    =
    \frac{
        |\Uparrow\rangle_{\rm I}
        +
        |\Downarrow\rangle_{\rm I}
    }{\sqrt2},
    \qquad
    H_1|\Downarrow\rangle_{\rm III}
    =
    \frac{
        |\Uparrow\rangle_{\rm I}
        -
        |\Downarrow\rangle_{\rm I}
    }{\sqrt2}.
\end{equation}
Thus, the logical $z$-basis of Case~III is mapped by the local
Clifford transformation $H_1$ onto the logical $x$-basis of Case~I.
After the corresponding relabeling of the logical axes, the logical
basis can be redefined to recover the $z$-basis displayed for
Case~I. The particular form of the logical basis in the canonical
representation therefore depends on which logical Pauli operator has
a local physical realization.

\section{Entanglement of Rank-2 Mixed Two-Qubit States}

We now consider the entanglement of rank-2 states {belonging to}
the two-dimensional logical subspaces introduced above. For a two-qubit
mixed state, the concurrence is defined by the convex-roof
construction {\cite{Wootters1998}}
\begin{equation}
    C(\rho)
    =
    \min_{\{p_i,|\psi_i\rangle\}}
    \sum_i
    p_i C(|\psi_i\rangle),
    \label{C(rho)}
\end{equation}
where the minimization is taken over all pure-state decompositions of
$\rho$.

For a pure two-qubit state,
\begin{equation}
    C(|\psi_i\rangle)
    =
    \sqrt{
        2
        \left(
            1-\Tr\rho_{1,i}^2
        \right)
    },
    \label{eq:concurrence_purity_pure}
\end{equation}
where
\begin{equation}
    \rho_{1,i}
    =
    \Tr_2
    |\psi_i\rangle\langle\psi_i|
\end{equation}
is the reduced density matrix of the first physical qubit.

We first consider Case~III of Table~\ref{tab:logical_operators}.
For a pure component
$|\psi_i\rangle=a_i|\Uparrow\rangle+b_i|\Downarrow\rangle$
of the decomposition of $\rho$, the physical Bloch vector
$\vec r_i$ of the first qubit is related to the corresponding logical
Bloch vector $\vec R_i$ by
\begin{align}
    r_i^x
    &=
    \langle\sigma_1^x\rangle_i
    =
    R_i^x\cos\theta,
    \\
    r_i^y
    &=
    \langle\sigma_1^y\rangle_i
    =
    R_i^y\cos\theta,
    \\
    r_i^z
    &=
    \langle\sigma_1^z\rangle_i
    =
    R_i^z.
\end{align}
Hence
\begin{equation}
    \vec r_i
    =
    \left(
        R_i^x\cos\theta,
        R_i^y\cos\theta,
        R_i^z
    \right).
    \label{r_i}
\end{equation}
This relation provides a simple geometrical interpretation of the
encoding. The logical Bloch sphere is mapped onto an ellipsoid in
the physical Bloch-vector space of the first qubit. The longitudinal
component is unchanged, while the two transverse components are
{scaled by the factor $\cos\theta$}.

Since $|\vec R_i|=1$ for a pure logical state,
\begin{align}
    |\vec r_i|^2
    &=
    \cos^2\theta
    \left[
        (R_i^x)^2+(R_i^y)^2
    \right]
    +(R_i^z)^2
    \nonumber\\
    &=
    \cos^2\theta
    +
    (R_i^z)^2\sin^2\theta.
\end{align}
Using Eq.~\eqref{eq:concurrence_purity_pure}, we obtain
\begin{equation}
    C(|\psi_i\rangle)
    =
    \sin\theta
    \sqrt{
        (R_i^x)^2+(R_i^y)^2
    }.
    \label{C(psi)}
\end{equation}

The same geometrical structure makes it possible to evaluate the
convex roof exactly for an arbitrary rank-2 mixed state
{in this logical subspace}. Substitution of
Eq.~\eqref{C(psi)} into Eq.~\eqref{C(rho)} gives
\begin{equation}
    C(\rho)
    =
    \sin\theta
    \min_{\{p_i,|\psi_i\rangle\}}
    \sum_i p_i
    \sqrt{
        (R_i^x)^2+(R_i^y)^2
    }.
\end{equation}

{Introducing the transverse logical Bloch vectors}
\begin{equation}
    \vec R_i^\perp=(R_i^x,R_i^y),
    \qquad
    \vec R^\perp=(R^x,R^y)
    =
    \sum_i p_i\vec R_i^\perp,
\end{equation}
{the triangle inequality immediately gives}
\begin{equation}
\sum_i p_i
|\vec R_i^\perp|
\geq
|\vec R^\perp|
=
\sqrt{(R^x)^2+(R^y)^2}.
\label{eq:lower_bound}
\end{equation}

{This lower bound can always be saturated. Let}
\begin{equation}
R_\perp
=
|\vec R^\perp|
=
\sqrt{(R^x)^2+(R^y)^2}.
\end{equation}
{For $R_\perp<1$, consider the two pure logical states
with Bloch vectors}
\begin{equation}
\vec R_\pm
=
\left(
R^x,
R^y,
\pm\sqrt{1-R_\perp^2}
\right).
\end{equation}
{Since $|\vec R|\leq1$, the longitudinal component
satisfies $|R^z|\leq\sqrt{1-R_\perp^2}$ and can therefore be reproduced
by choosing}
\begin{equation}
p_\pm
=
\frac{1}{2}
\left(
1
\pm
\frac{R^z}{\sqrt{1-R_\perp^2}}
\right).
\end{equation}
{Indeed,
$\vec R=p_+\vec R_++p_-\vec R_-$, while both pure states have the same
transverse magnitude $R_\perp$. Hence}
\begin{equation}
\sum_{\nu=\pm}
p_\nu|\vec R_\nu^\perp|
=
R_\perp,
\end{equation}
{which saturates Eq.~\eqref{eq:lower_bound}. The limiting
case $R_\perp=1$ corresponds to $R^z=0$ and $|\vec R|=1$, i.e., to a
pure logical state, for which the result follows directly.}

The concurrence of the considered rank-2 family therefore takes the
closed form
\begin{equation}
C(\rho)
=
\sin\theta
\sqrt{
(R^x)^2+(R^y)^2
}.
\label{eq:concurrence_final}
\end{equation}

The parameter $\theta$, which determines the deformation of the
physical Bloch sphere, can itself be obtained directly from a local
measurement. For all states belonging to the Case~III logical subspace,
\begin{equation}
\langle I_1\otimes\sigma_2^z\rangle
=
\cos\theta,
\end{equation}
and hence
\begin{equation}
\sin\theta
=
\sqrt{
1-
\langle I_1\otimes\sigma_2^z\rangle^2
}.
\end{equation}
Using
\begin{equation}
R^x=\langle\sigma_1^x\otimes\sigma_2^z\rangle,
\qquad
R^y=\langle\sigma_1^y\otimes\sigma_2^z\rangle,
\end{equation}
the concurrence can therefore be reconstructed entirely from directly
measurable physical observables:
\begin{equation}
C(\rho)
=
\sqrt{
1-
\langle I_1\otimes\sigma_2^z\rangle^2
}
\sqrt{
\langle\sigma_1^x\otimes\sigma_2^z\rangle^2
+
\langle\sigma_1^y\otimes\sigma_2^z\rangle^2
}.
\label{eq:C_caseIII}
\end{equation}

{Importantly,
$\langle I_1\otimes\sigma_2^z\rangle$ is obtained as a marginal of
either of the two correlation measurements above. Thus, no additional
measurement setting is required to determine the factor
$\sin\theta$, and the concurrence in Case~III can be reconstructed
from only two measurement settings.}

The same construction applies to the other two canonical encodings
listed in Table~\ref{tab:logical_operators}. Thus,
\begin{equation}
C(\rho)
=
\sqrt{
    1-\langle I_1\otimes\sigma_2^z\rangle^2
}
\begin{cases}
\sqrt{
    \langle\sigma_1^y\otimes\sigma_2^z\rangle^2
    +
    \langle\sigma_1^z\otimes\sigma_2^z\rangle^2
},
& \text{Case I}, \\

\sqrt{
    \langle\sigma_1^x\otimes\sigma_2^z\rangle^2
    +
    \langle\sigma_1^z\otimes\sigma_2^z\rangle^2
},
& \text{Case II}, \\

\sqrt{
    \langle\sigma_1^x\otimes\sigma_2^z\rangle^2
    +
    \langle\sigma_1^y\otimes\sigma_2^z\rangle^2
},
& \text{Case III}.
\end{cases}
\label{eq:concurrence_rank2_all}
\end{equation}

{In each case, the local expectation value
$\langle I_1\otimes\sigma_2^z\rangle$ can be extracted from the same
measurement data used for the two correlations entering the
corresponding expression. Hence all three canonical realizations
require only two measurement settings.}

It is useful to compare Eq.~\eqref{eq:concurrence_final} with earlier
results for particular rank-2 subspaces. In
Refs.~\cite{Frydryszak2017,Kuzmak2021}, states belonging to subspaces
such as
$\{|00\rangle,|11\rangle\}$ or
$\{|01\rangle,|10\rangle\}$ were considered, and the corresponding
concurrence takes the form
\begin{equation}
C(\rho)
=
\sqrt{
\langle\Sigma^x\rangle^2
+
\langle\Sigma^y\rangle^2
}.
\label{eq:previous_concurrence}
\end{equation}

These subspaces are recovered from the present construction for
$\cos\theta=0$ ($\sin\theta=1$), up to local basis transformations,
and Eq.~\eqref{eq:concurrence_final} reduces directly to
Eq.~\eqref{eq:previous_concurrence}.

The present analysis extends this result to arbitrary $\theta$ and,
within the class of two-qubit encodings generated by factorizable
logical Pauli operators, covers all nontrivial canonical realizations
identified in Sec.~3. The factor
$\sin\theta=\sqrt{1-\cos^2\theta}$ accounts for
{the distinguishability of the complementary-qubit
states, equivalently determining} the transverse deformation of the
physical Bloch sphere.

\section{{Generalization to Bipartite $N$-Qubit States}}

The two-qubit construction can be extended to an $N$-qubit system
partitioned into a subsystem $A$ containing $k$ qubits and a
complementary subsystem $B$ containing the remaining $N-k$ qubits,
with $1\leq k<N$. We restrict the discussion to the generalization
of Case~III; the other canonical encodings can be treated analogously.

The logical Pauli operators are chosen as
\begin{align}
    \Sigma^x
    &=
    \left(
        \bigotimes_{i=1}^{k}\sigma_i^x
    \right)
    \otimes
    \left(
        \bigotimes_{j=k+1}^{N}\sigma_j^z
    \right),\\
    \Sigma^y
    &=
    \sigma_1^y
    \otimes
    \left(
        \bigotimes_{i=2}^{k}\sigma_i^x
    \right)
    \otimes
    \left(
        \bigotimes_{j=k+1}^{N}\sigma_j^z
    \right),\\
    \Sigma^z
    &=
    \sigma_1^z\otimes I_{N-1}.
    \label{eq:bipartite_operators}
\end{align}

A logical basis consistent with these operators is given by
\begin{align}
|\Uparrow\rangle
&=
|00\ldots0\rangle_A
\otimes
\bigotimes_{j=k+1}^{N}
\left(
    \cos\frac{\theta_j}{2}|0\rangle_j
    +
    e^{i\phi_j}\sin\frac{\theta_j}{2}|1\rangle_j
\right),
\label{eq:bipartite_up}
\\
|\Downarrow\rangle
&=
|11\ldots1\rangle_A
\otimes
\bigotimes_{j=k+1}^{N}
\left(
    \cos\frac{\theta_j}{2}|0\rangle_j
    -
    e^{i\phi_j}\sin\frac{\theta_j}{2}|1\rangle_j
\right).
\label{eq:bipartite_down}
\end{align}

For the bipartition $A|B$, we use the generalized concurrence
(also known as the $I$-concurrence) \cite{Rungta2001}. For a
pure bipartite state it is defined as
\begin{equation}
    C(|\psi_i\rangle)
    =
    \sqrt{
        2\left(
            1-\Tr\rho_{A,i}^2
        \right)
    },
    \qquad
    \rho_{A,i}
    =
    \Tr_B|\psi_i\rangle\langle\psi_i|,
    \label{eq:generalized_concurrence_pure}
\end{equation}
and for mixed states it is extended through the convex-roof
construction,
\begin{equation}
    C(\rho)
    =
    \min_{\{p_i,|\psi_i\rangle\}}
    \sum_i p_i C(|\psi_i\rangle).
    \label{eq:generalized_concurrence_mixed}
\end{equation}

Although the Hilbert spaces of $A$ and $B$ may have dimensions
larger than two, the states considered here involve only the two
states $|00\ldots0\rangle_A$ and $|11\ldots1\rangle_A$ of subsystem
$A$. Hence their Schmidt rank
across the $A|B$ bipartition is at most two. For $N=2$ and $k=1$,
this construction reduces to the usual two-qubit case.

For a pure logical state
$|\psi_i\rangle=a_i|\Uparrow\rangle+b_i|\Downarrow\rangle$, denote
the corresponding states of subsystem $B$ by
$|\chi_+\rangle_B$ and $|\chi_-\rangle_B$. Their overlap factorizes as
\begin{equation}
\eta
=
\langle\chi_-|\chi_+\rangle
=
\prod_{j=k+1}^{N}
\left(
\cos^2\frac{\theta_j}{2}
-
\sin^2\frac{\theta_j}{2}
\right)
=
\prod_{j=k+1}^{N}\cos\theta_j.
\label{eq:eta_overlap}
\end{equation}
This collective overlap plays the same role as $\cos\theta$ in the
two-qubit case.

After tracing out subsystem $B$, the reduced state of subsystem $A$
is completely described within {this effective
two-dimensional subspace}. Its Bloch vector is
\begin{equation}
    \vec r_i
    =
    \left(
        \eta R_i^x,
        \eta R_i^y,
        R_i^z
    \right),
\end{equation}
where $R_i^\alpha=\langle\Sigma^\alpha\rangle_i$ are the components
of the logical Bloch vector. Thus, the complementary subsystem
produces the same transverse deformation of the effective Bloch
sphere as in the two-qubit case, now characterized by the collective
factor $\eta$.

Since $|\vec R_i|=1$ for a pure logical state,
\begin{equation}
    |\vec r_i|^2
    =
    \eta^2
    +
    (1-\eta^2)(R_i^z)^2.
\end{equation}
Substitution into Eq.~\eqref{eq:generalized_concurrence_pure} gives
\begin{equation}
    C(|\psi_i\rangle)
    =
    \sqrt{1-\eta^2}
    \sqrt{
        (R_i^x)^2+(R_i^y)^2
    }.
    \label{eq:C_pure_squashed_bipartite}
\end{equation}

For a rank-2 mixed state {belonging to} the same logical
subspace, the convex-roof minimization proceeds exactly as in the
two-qubit case. Therefore,
\begin{equation}
    C(\rho)
    =
    \sqrt{1-\eta^2}
    \sqrt{
        (R^x)^2+(R^y)^2
    }.
    \label{eq:concurrence_N_qubit_bipartite}
\end{equation}

Importantly, $\eta$ does not have to be determined from the
state-preparation parameters. It can be measured directly as the
multi-qubit correlation of subsystem $B$,
\begin{equation}
    \eta
    =
    \left\langle
        \bigotimes_{j=k+1}^{N}\sigma_j^z
    \right\rangle.
    \label{eq:eta_measurement}
\end{equation}
Together with
\begin{align}
    R^x
    =
    \left\langle
        \left(
            \bigotimes_{i=1}^{k}\sigma_i^x
        \right)
        \otimes
        \left(
            \bigotimes_{j=k+1}^{N}\sigma_j^z
        \right)
    \right\rangle,\ \ \ 
    R^y
    =
    \left\langle
        \sigma_1^y
        \otimes
        \left(
            \bigotimes_{i=2}^{k}\sigma_i^x
        \right)
        \otimes
        \left(
            \bigotimes_{j=k+1}^{N}\sigma_j^z
        \right)
    \right\rangle,
\end{align}
this allows the bipartite concurrence to be reconstructed directly
from {two multi-qubit Pauli measurement settings}.
{Indeed, the correlation in
Eq.~\eqref{eq:eta_measurement} involves only the $Z$ measurements on
subsystem $B$ and can be obtained as a marginal from either of the
same measurement settings used to determine $R^x$ or $R^y$.}

For $N=2$ and $k=1$, $\eta=\cos\theta$, and
Eq.~\eqref{eq:concurrence_N_qubit_bipartite} reduces to
\begin{equation}
    C(\rho)
    =
    \sin\theta
    \sqrt{
        (R^x)^2+(R^y)^2
    },
\end{equation}
recovering the two-qubit result.

For the rank-2 family considered here, the concurrence can also be
related directly to the global and subsystem purities. The purity of
the full state is
\begin{equation}
    \Tr\rho^2
    =
    \frac{1}{2}
    \left[
        1+(R^x)^2+(R^y)^2+(R^z)^2
    \right],
\end{equation}
whereas the reduced state of subsystem $A$ has purity
\begin{equation}
    \Tr\rho_A^2
    =
    \frac{1}{2}
    \left[
        1+
        \eta^2
        \left(
            (R^x)^2+(R^y)^2
        \right)
        +(R^z)^2
    \right].
\end{equation}
Using Eq.~\eqref{eq:concurrence_N_qubit_bipartite}, their difference
gives
\begin{equation}
    C^2(\rho)
    =
    2
    \left(
        \Tr\rho^2-\Tr\rho_A^2
    \right).
    \label{eq:purity_concurrence_relation}
\end{equation}
This identity is specific to the rank-2 logical family considered
here and does not hold for arbitrary mixed bipartite states. It also
provides an alternative way to access the concurrence through
measurements of the global and subsystem purities, for example by
two-copy interference or SWAP-type protocols.

\section{Numerical Simulations and Error Mitigation}

To validate the analytical results for the concurrence of rank-2 mixed
states, we implement the state-preparation and correlation-measurement
protocol using the noise model of the IBM Quantum
\texttt{FakeKingston} backend. Rather than performing full quantum
state tomography, the concurrence is reconstructed from a small set
of logical Pauli correlations.
{
For the two-qubit system, the noisy simulation is compared directly
with the exact result. For the five-qubit register, readout-error
mitigation is applied, while for the ten-qubit register it is
supplemented by zero-noise extrapolation (ZNE) to compensate for the
increased circuit noise.
}

\subsection{Two-qubit case}

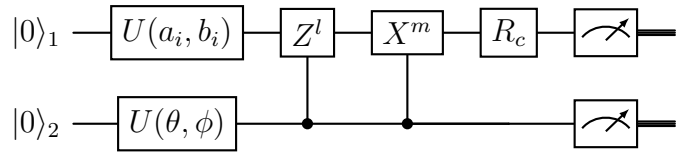
\begin{figure}[htbp]
\centering
\begin{quantikz}
\lstick{$|0\rangle_1$}
& \gate{U(a_i,b_i)}
& \gate{Z^l}
& \gate{X^m}
& \gate{R_c}
& \meter{}
& \cw\\
\lstick{$|0\rangle_2$}
& \gate{U(\theta,\phi)}
& \ctrl{-1}
& \ctrl{-1}
& \qw
& \meter{}
& \cw
\end{quantikz}
\caption{Unified circuit for preparation and measurement of the
two-qubit logical states. The binary parameters $l,m\in\{0,1\}$
select Case~I ($l=0,m=1$), Case~II ($l=1,m=1$), or Case~III
($l=1,m=0$). The rotation
$R_c\in\{I,R_x(\pi/2),R_y(-\pi/2)\}$ selects the measurement basis.}
\label{fig:prep_and_measurement_unified}
\end{figure}

An arbitrary pure logical state
$
    |\psi_i\rangle
    =
    a_i|\Uparrow\rangle+b_i|\Downarrow\rangle,
    i\in\{1,2\},
$
is prepared from the computational state $|00\rangle$ using the
single-qubit rotations
\begin{align}
    U(a_i,b_i)|0\rangle
    &=
    a_i|0\rangle+b_i|1\rangle,\\
    U(\theta,\phi)|0\rangle
    &=
    \cos\frac{\theta}{2}|0\rangle
    +
    e^{i\phi}\sin\frac{\theta}{2}|1\rangle,
\end{align}
followed by conditional two-qubit operations, as shown in
Fig.~\ref{fig:prep_and_measurement_unified}.

The three choices of $(l,m)$ reproduce the canonical encodings listed
in Table~\ref{tab:logical_operators}. Since they are locally
Clifford-equivalent, in the following numerical analysis we focus on
Case~III,
\begin{equation}
    \Sigma^x=X_1Z_2,
    \qquad
    \Sigma^y=Y_1Z_2,
    \qquad
    \Sigma^z=Z_1I_2.
\end{equation}

To construct the mixed state
\begin{equation}
    \rho(p)
    =
    p|\psi_1\rangle\langle\psi_1|
    +(1-p)|\psi_2\rangle\langle\psi_2|,
\end{equation}
the two pure component states are executed independently.
{
The expectation values of the mixture are then reconstructed by linear
combination of the measured observables.
}
Each circuit is sampled with
$N_{\rm shots}=1024$. For each value of $\theta$, two component
states and two measurement settings are used, corresponding to
$4096$ shots in total.

We consider two representative pairs of logical states:
\begin{align}
\text{Set A:}\qquad
&|\psi_1\rangle=|\bplus\rangle,\quad
|\psi_2\rangle=|\bplusi\rangle,
\\
\text{Set B:}\qquad
&|\psi_1\rangle=|\bplus\rangle, \quad
|\psi_2\rangle=|\bminus\rangle,
\end{align}
where
\begin{equation}
|\bplus\rangle=
\frac{|\Uparrow\rangle+|\Downarrow\rangle}{\sqrt2},
\qquad
|\bplusi\rangle=
\frac{|\Uparrow\rangle+i|\Downarrow\rangle}{\sqrt2},
\qquad
|\bminus\rangle=
\frac{|\Uparrow\rangle-|\Downarrow\rangle}{\sqrt2}.
\end{equation}
Set~A generates both transverse components $R^x$ and $R^y$, whereas
for Set~B the states are antipodal along the logical $x$ axis and
$R^y=0$.

To measure $\Sigma^x=X_1Z_2$ and $\Sigma^y=Y_1Z_2$, the rotations
$R_y(-\pi/2)$ and $R_x(\pi/2)$, respectively, are applied to the
first qubit before computational-basis measurement. The expectation
values are obtained from the measured parity as
\begin{equation}
    \langle\Sigma^{x,y}\rangle_i
    =
    \frac{
        N_{00}+N_{11}-N_{01}-N_{10}
    }{N_{\rm shots}},
\end{equation}
where $N_{b_0b_1}$ denotes the number of outcomes with bit string
$b_0b_1$.

The same measurement data also provide the marginal correlation
\begin{equation}
    \eta_i
    =
    \langle I_1\otimes Z_2\rangle_i
    =
    \frac{
        N_{00}+N_{10}-N_{01}-N_{11}
    }{N_{\rm shots}},
\end{equation}
which ideally equals $\cos\theta$.
{
The estimates obtained from the two measurement settings are averaged,
so no additional circuit is required to determine the factor
$\sqrt{1-\eta^2}$ entering the concurrence.
}

\begin{figure}[htbp]
\centering
\begin{subfigure}[b]{0.48\linewidth}
    \centering
    \includegraphics[width=\linewidth]{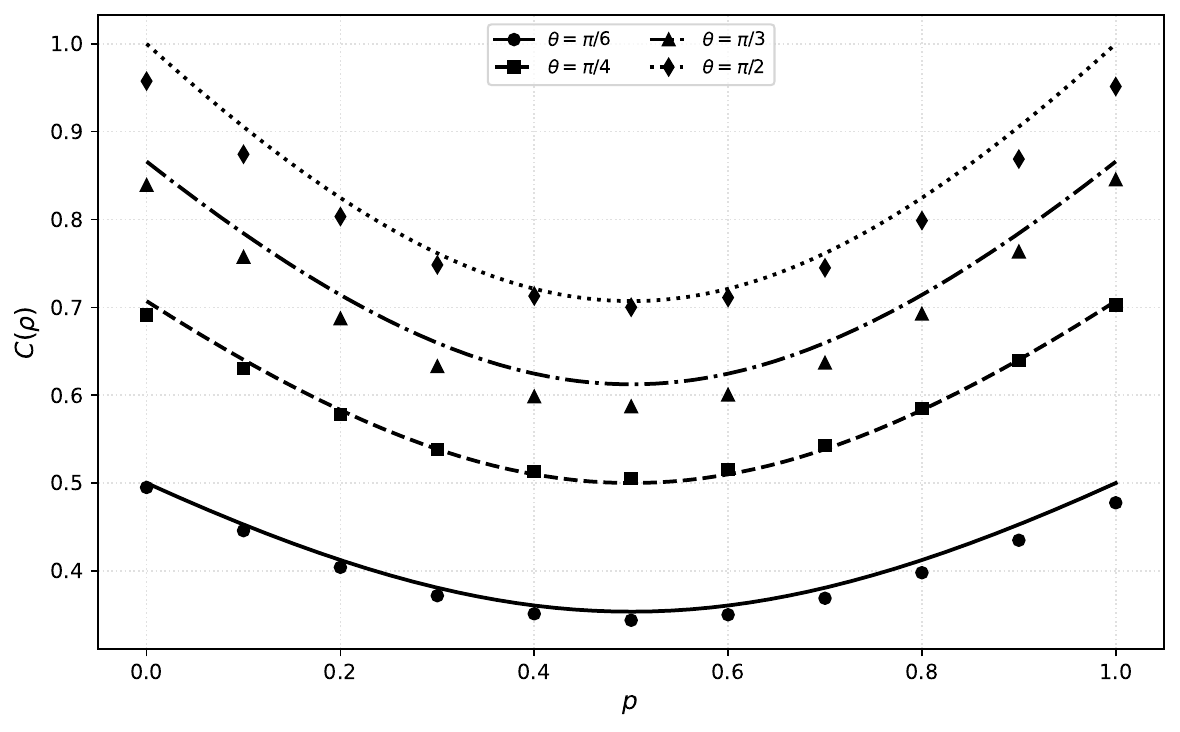}
    \caption{Set A ($|\bplus\rangle,|\bplusi\rangle$)}
    \label{fig:conc_setA}
\end{subfigure}
\hfill
\begin{subfigure}[b]{0.48\linewidth}
    \centering
    \includegraphics[width=\linewidth]{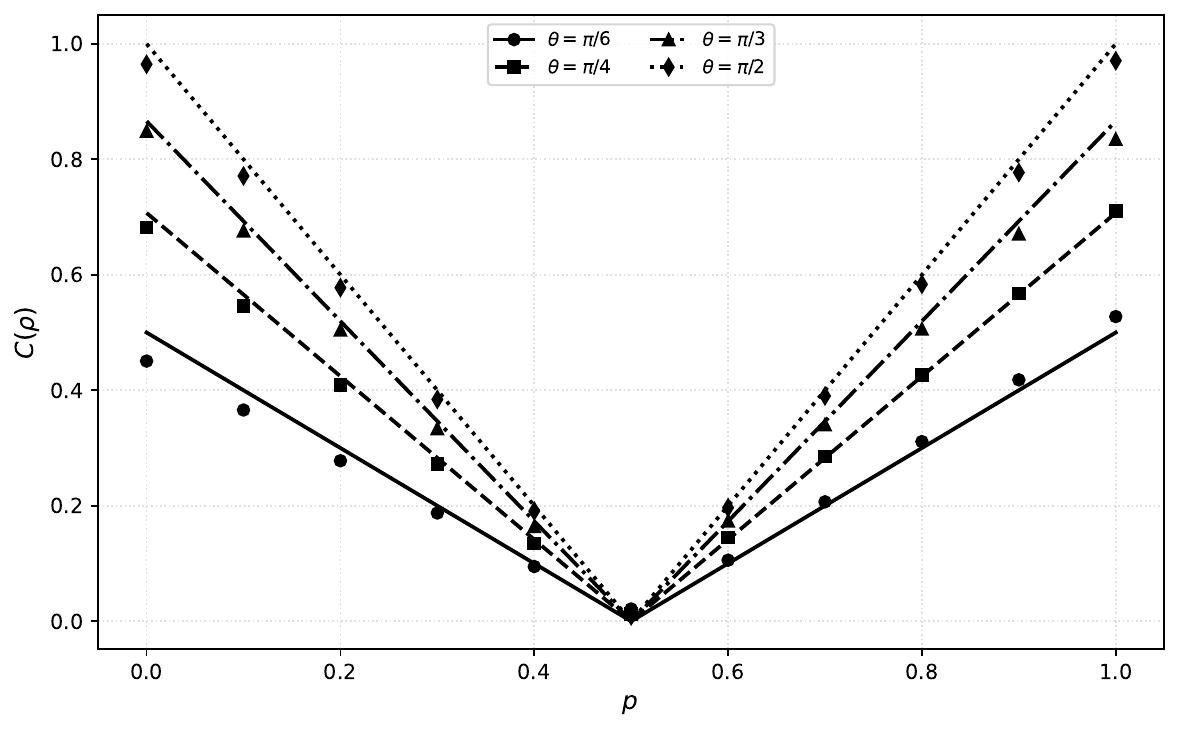}
    \caption{Set B ($|\bplus\rangle,|\bminus\rangle$)}
    \label{fig:conc_setB}
\end{subfigure}
\caption{Reconstructed concurrence $C(\rho)$ as a function of the
mixing parameter $p$ for
$\theta\in\{\pi/6,\pi/4,\pi/3,\pi/2\}$.
{
Lines show the exact results, while filled symbols represent noisy
simulations using the \texttt{FakeKingston} backend model.
Different line styles and marker shapes correspond to different
values of $\theta$.
}}
\label{fig:concurrence_results}
\end{figure}

For a mixture with parameter $p$, linearity of expectation values
gives
\begin{equation}
    R^{x,y}_{\rm sim}
    =
    pR^{x,y}_1+(1-p)R^{x,y}_2,
    \qquad
    \eta_{\rm sim}
    =
    p\eta_1+(1-p)\eta_2.
\end{equation}
The reconstructed concurrence is then
\begin{equation}
    C_{\rm sim}(\rho)
    =
    \sqrt{1-\eta_{\rm sim}^2}
    \sqrt{
        (R^x_{\rm sim})^2+(R^y_{\rm sim})^2
    }.
\end{equation}

As shown in Fig.~\ref{fig:concurrence_results}, the noisy simulations
closely reproduce the analytical dependence of the concurrence on
$p$ and $\theta$. For Set~B, the transverse Bloch vectors of the two
component states are opposite, and their contributions cancel at
$p=1/2$, where the concurrence vanishes for all values of $\theta$.
\subsection{Multi-qubit case}

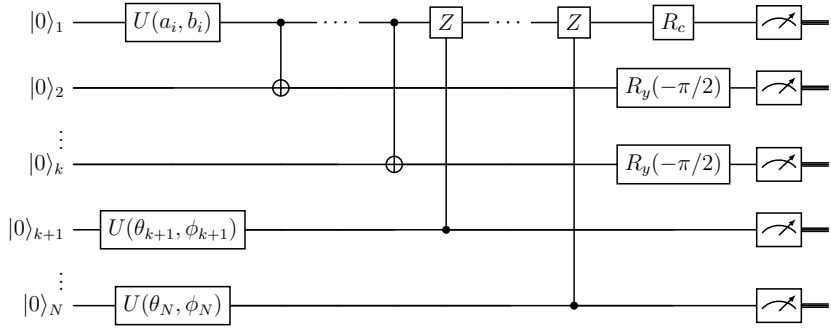
\begin{figure}[htbp]
\centering
\resizebox{0.65\textwidth}{!}{%
\begin{quantikz}
\lstick{$|0\rangle_1$}
& \gate{U(a_i,b_i)}
& \ctrl{1}
& \gate[style={draw=none}]{\cdots}
& \ctrl{3}
& \gate{Z}
& \gate[style={draw=none}]{\cdots}
& \gate{Z}
& \gate{R_c}
& \meter{}
& \cw
\\
\lstick{$|0\rangle_2$}
& \qw
& \targ{}
& \qw
& \qw
& \qw
& \qw
& \qw
& \gate{R_y(-\pi/2)}
& \meter{}
& \cw
\\
\lstick{$\vdots$}
\\[-0.3cm]
\lstick{$|0\rangle_k$}
& \qw
& \qw
& \qw
& \targ{}
& \qw
& \qw
& \qw
& \gate{R_y(-\pi/2)}
& \meter{}
& \cw
\\
\lstick{$|0\rangle_{k+1}$}
& \gate{U(\theta_{k+1},\phi_{k+1})}
& \qw
& \qw
& \qw
& \ctrl{-4}
& \qw
& \qw
& \qw
& \meter{}
& \cw
\\
\lstick{$\vdots$}
\\[-0.3cm]
\lstick{$|0\rangle_N$}
& \gate{U(\theta_N,\phi_N)}
& \qw
& \qw
& \qw
& \qw
& \qw
& \ctrl{-6}
& \qw
& \meter{}
& \cw
\end{quantikz}}

\caption{Schematic $N$-qubit state-preparation and measurement circuit.
The first qubit is prepared by $U(a_i,b_i)$ and entangled with the
remaining qubits of subsystem $A$ by CNOT gates. The qubits of
subsystem $B$ are prepared by $U(\theta_j,\phi_j)$ and coupled to the
first qubit through controlled-$Z$ gates. For the measurement of
$R^x$, $R_c=R_y(-\pi/2)$, while for $R^y$,
$R_c=R_x(\pi/2)$. The remaining qubits of subsystem $A$ are rotated
by $R_y(-\pi/2)$, and the qubits of subsystem $B$ are measured
directly in the computational basis.}
\label{fig:prep_and_measurement_n_qubit}
\end{figure}

The protocol extends directly to an $N$-qubit bipartition
$A|B$, with $k$ qubits in $A$ and $N-k$ qubits in $B$.
The corresponding state-preparation and measurement circuit is shown
in Fig.~\ref{fig:prep_and_measurement_n_qubit}.

For the generalized logical operators of
Eq.~\eqref{eq:bipartite_operators}, the required transverse
correlations are
\begin{align}
R^x=
\left\langle
X_1\cdots X_k Z_{k+1}\cdots Z_N
\right\rangle,\ \ \
R^y=
\left\langle
Y_1X_2\cdots X_k Z_{k+1}\cdots Z_N
\right\rangle.
\end{align}
Thus, $R^x$ is measured by rotating all qubits of $A$ to the
$x$ basis, whereas for $R^y$ the first qubit is rotated to the
$y$ basis and the remaining $k-1$ qubits to the $x$ basis.
Subsystem $B$ is measured directly in the computational basis.

The overlap factor is extracted from the same outcomes,
\begin{equation}
    \eta_{\rm sim}
    =
    \left\langle
        Z_{k+1}\cdots Z_N
    \right\rangle,
\end{equation}
giving
\begin{equation}
    C_{\rm sim}(\rho)
    =
    \sqrt{1-\eta_{\rm sim}^2}
    \sqrt{(R^x_{\rm sim})^2+(R^y_{\rm sim})^2}.
\end{equation}

We consider $N=5$ with $k=1,3$ and $N=10$ with $k=1$,
using the mixed states
\begin{equation}
    |\psi_1\rangle=|\bplus\rangle,
    \qquad
    |\psi_2\rangle=|\bminusi\rangle.
\end{equation}
For $N=5$, the uniform configurations
$\theta_j=\pi/6$ and $\theta_j=\pi/4$ are supplemented by
\begin{equation}
\boldsymbol{\theta}^{(5,1)}_{\rm mix}
=
\left(\frac{\pi}{6},\frac{\pi}{4},
\frac{\pi}{3},\frac{\pi}{2}\right),
\qquad
\boldsymbol{\theta}^{(5,3)}_{\rm mix}
=
\left(\frac{\pi}{6},\frac{\pi}{3}\right).
\end{equation}
For $N=10$, $k=1$, we compare $\theta_j=\pi/6$ with
\begin{equation}
\boldsymbol{\theta}^{(10,1)}_{\rm mix}
=
\left(
\frac{\pi}{6},\frac{\pi}{4},\frac{\pi}{3},\frac{\pi}{2},
\frac{\pi}{6},\frac{\pi}{4},\frac{\pi}{3},\frac{\pi}{2},
\frac{\pi}{6}
\right),
\end{equation}
with $\phi_j=\pi/2$ in all cases.
\begin{figure}[htbp]
    \centering
    \begin{subfigure}[b]{0.32\linewidth}
        \centering
        \includegraphics[width=\linewidth]
        {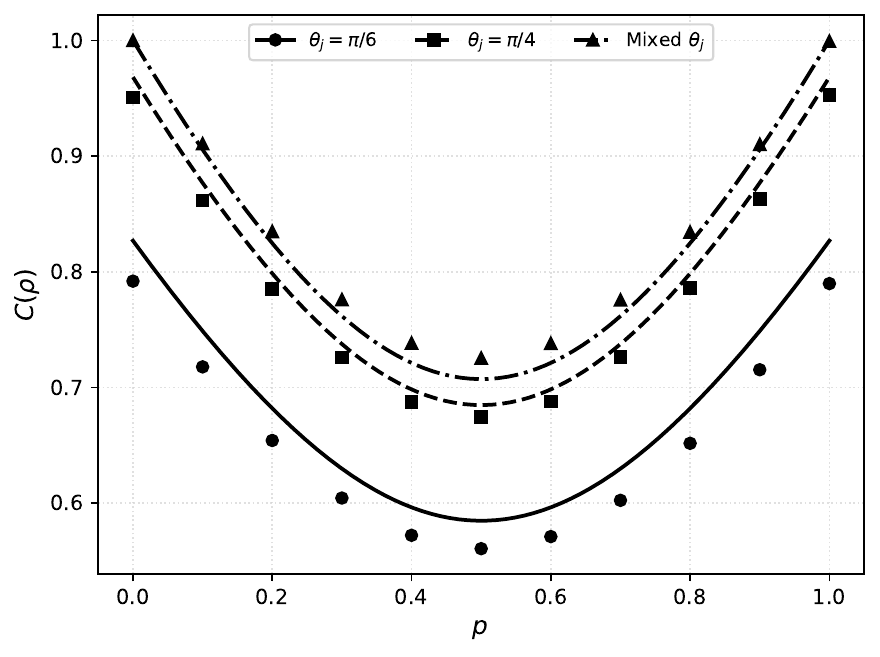}
        \caption{$N=5,\ k=1$}
        \label{fig:conc_n5_k1_all}
    \end{subfigure}
    \hfill
    \begin{subfigure}[b]{0.32\linewidth}
        \centering
        \includegraphics[width=\linewidth]
        {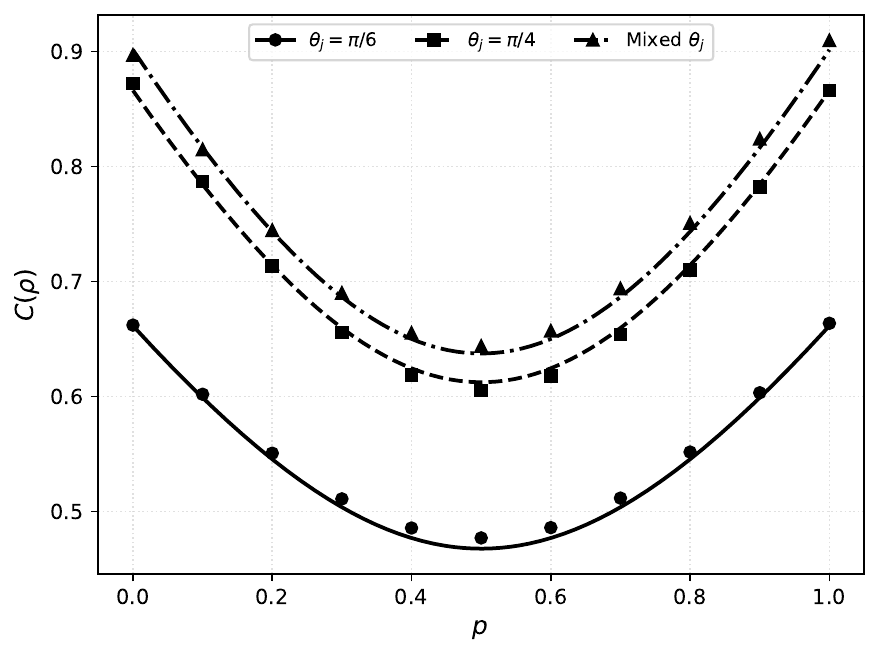}
        \caption{$N=5,\ k=3$}
        \label{fig:conc_n5_k3_all}
    \end{subfigure}
    \hfill
    \begin{subfigure}[b]{0.32\linewidth}
        \centering
        \includegraphics[width=\linewidth]
        {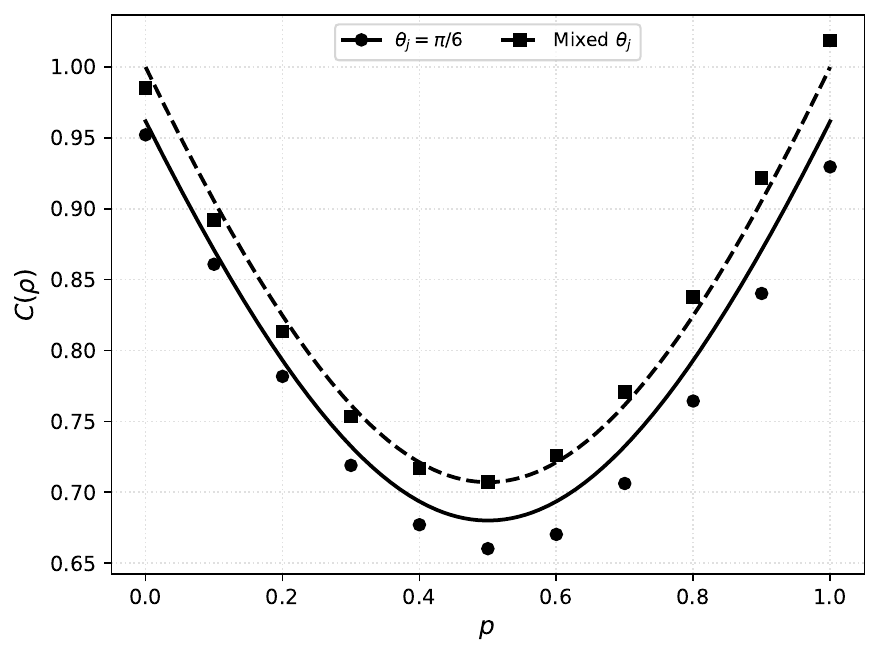}
        \caption{$N=10,\ k=1$}
        \label{fig:conc_n10_k1_all}
    \end{subfigure}

    \caption{Concurrence $C(\rho)$ as a function of $p$ for uniform
    and mixed angle configurations. Lines show the exact results and
    filled symbols the error-mitigated \texttt{FakeKingston}
    simulations. For $N=5$, readout mitigation is used, whereas for
    $N=10$ it is supplemented by linear ZNE.}
    \label{fig:concurrence_multi_qubit_all}
\end{figure}

For $N=5$, the circuits are transpiled at optimization level~3 and
only readout-error mitigation is applied. The latter uses the full
$2^N\times2^N$ assignment matrix $M_{\rm cal}$, which is employed
here as a numerical benchmark rather than as a scalable mitigation
scheme. For $N=10$, readout mitigation is supplemented by linear ZNE
using global unitary folding with $\lambda\in\{1,3,5\}.$ 
Each measurement and calibration circuit is
sampled with $N_{\rm shots}=4096$.

Figure~\ref{fig:concurrence_multi_qubit_all} shows good agreement
between the reconstructed and exact concurrence. Readout mitigation
provides a substantial correction for $N=5$, while for $N=10$ the
additional ZNE step further reduces the noise-induced deviation.

\begin{figure}[htbp]
    \centering
    \begin{subfigure}[b]{0.32\linewidth}
        \centering
        \includegraphics[width=\linewidth]
        {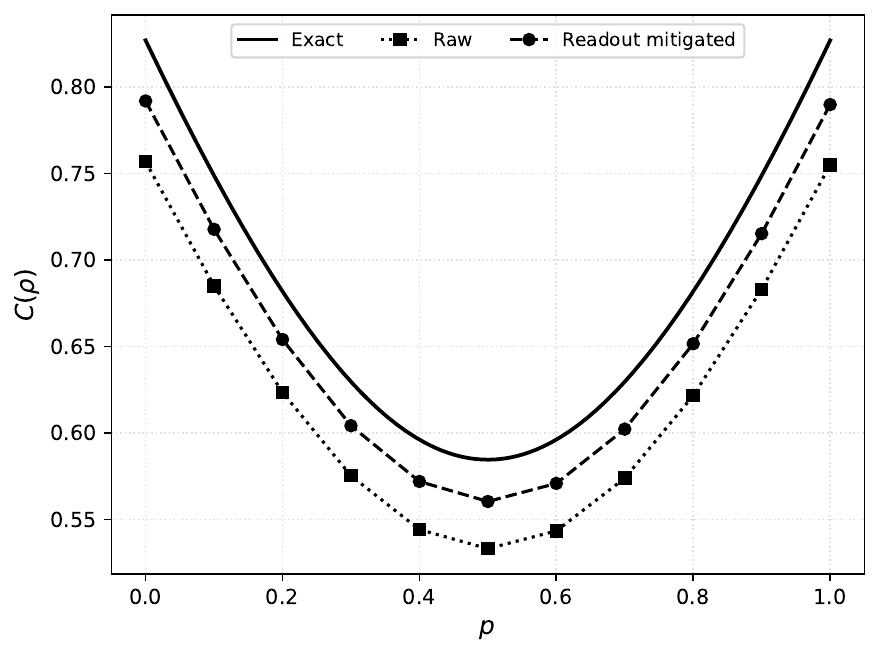}
        \caption{$N=5,\ k=1,\ \theta_j=\pi/6$}
        \label{fig:mit_n5_k1_pi6}
    \end{subfigure}
    \hfill
    \begin{subfigure}[b]{0.32\linewidth}
        \centering
        \includegraphics[width=\linewidth]
        {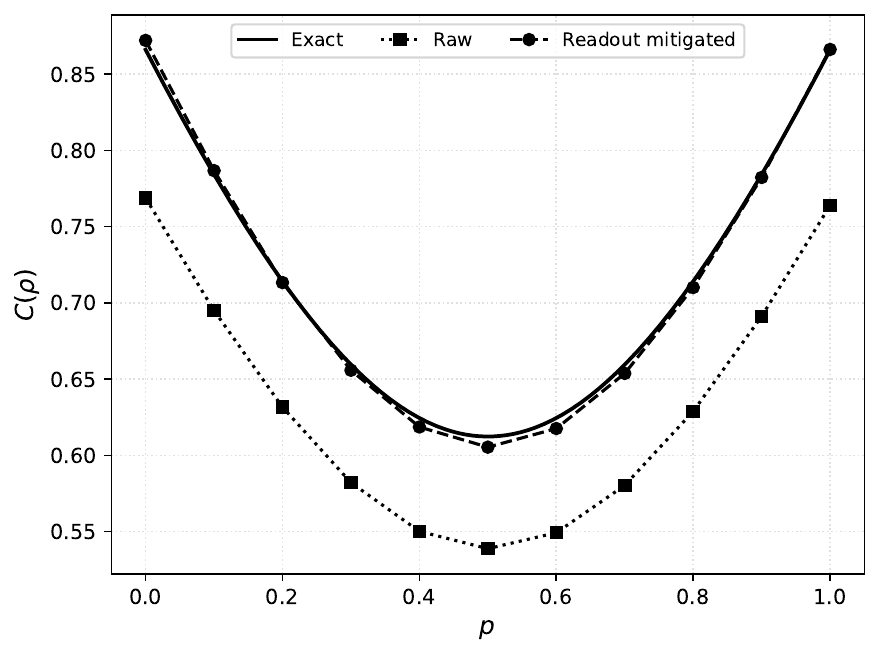}
        \caption{$N=5,\ k=3,\ \theta_j=\pi/4$}
        \label{fig:mit_n5_k3_pi4}
    \end{subfigure}
    \hfill
    \begin{subfigure}[b]{0.315\linewidth}
        \centering
        \includegraphics[width=\linewidth]
        {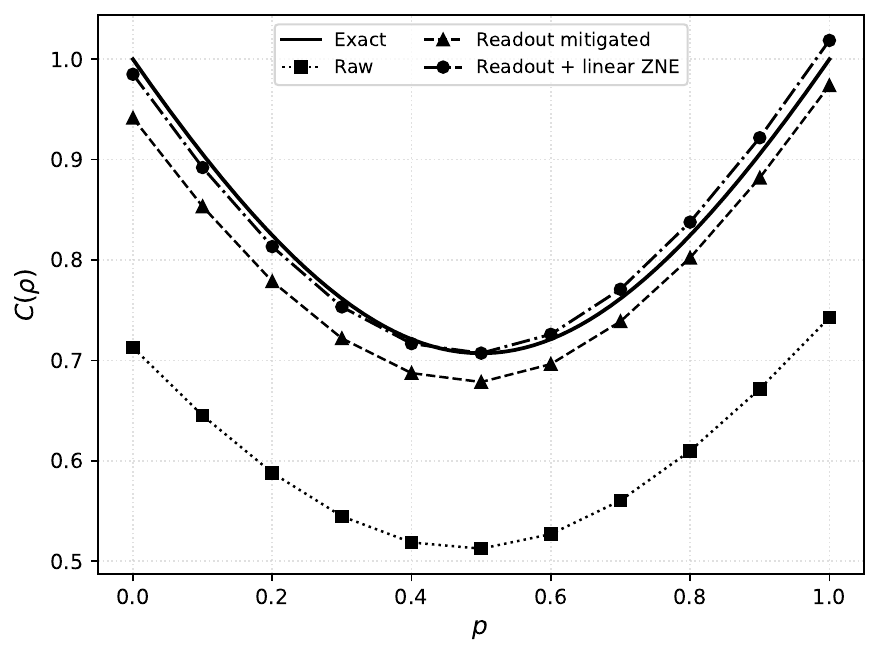}
        \caption{$N=10,\ k=1,\ 
        \boldsymbol{\theta}=\boldsymbol{\theta}^{(10,1)}_{\rm mix}$}
        \label{fig:mit_n10_k1_mixed}
    \end{subfigure}

    \caption{Representative comparison of the exact, raw, and
    mitigated concurrence. For $N=5$, only readout mitigation is
    applied, while for $N=10$ linear ZNE is additionally included.}
    \label{fig:mitigation_breakdown_comparative}
\end{figure}

Figure~\ref{fig:mitigation_breakdown_comparative} illustrates the
increasing effect of noise with register size. Readout correction is
sufficient to recover much of the signal for $N=5$, whereas the
$N=10$ case benefits from the additional ZNE correction.

\section{Conclusion}

We developed a correlation-based framework for quantifying the entanglement of rank-2 mixed states without full quantum state tomography. For two-qubit systems, we classified the nontrivial factorizable Pauli encodings and showed that the canonical realizations are locally Clifford-equivalent. Within the corresponding logical subspaces, the relation between logical and physical Bloch vectors has a simple geometrical interpretation in terms of a transversely deformed Bloch sphere.

This structure allows the convex-roof concurrence to be evaluated analytically. For two qubits, the concurrence is given by the transverse logical Bloch components multiplied by $\sin\theta$. For an arbitrary bipartition of an $N$-qubit system into $k$ and $N-k$ qubits, the same form is retained with the scaling factor
\begin{equation}
    \sqrt{1-\eta^2},
    \qquad
    \eta=\prod_{j=k+1}^{N}\cos\theta_j.
\end{equation}

Importantly, $\cos\theta$ and $\eta$ can be extracted directly from computational-basis correlations, so the concurrence is reconstructed from a small set of multi-qubit Pauli expectation values. For the considered rank-2 family, we also obtained
\begin{equation}
    C^2(\rho)
    =
    2\left(
        \Tr\rho^2-\Tr\rho_A^2
    \right),
\end{equation}
linking entanglement to global and subsystem purities.

The analytical results were validated by noisy simulations using the IBM Quantum \texttt{FakeKingston} backend model for systems of up to ten qubits. Readout mitigation, and for the ten-qubit case additionally zero-noise extrapolation, improved the agreement with the exact predictions. The proposed method therefore provides a low-measurement-overhead approach for estimating entanglement in the considered rank-2 states and extends naturally to multi-qubit logical registers.

\section*{Acknowledgements}

This work was supported by the National Research Foundation of Ukraine
under Project No.~2025.07/0108.

\end{document}